\documentstyle[12pt]{article}
\def\singlespace {\smallskipamount=3.75pt plus1pt minus1pt
                  \medskipamount=7.5pt plus2pt minus2pt
                  \bigskipamount=15pt plus4pt minus4pt
                  \normalbaselineskip=15pt plus0pt minus0pt
                  \normallineskip=1pt
                  \normallineskiplimit=0pt
                  \jot=3.75pt
                  {\def\smallskip {\vskip\smallskipamount}}
                  {\def\medskip   {\vskip\medskipamount}}
                  {\def\bigskip   {\vskip\bigskipamount}}
                  {\setbox\strutbox=\hbox{\vrule 
                    height10.5pt depth4.5pt width 0pt}}
                  \parskip 7.5pt
                  \normalbaselines}
\def\middlespace {\smallskipamount=5.625pt plus1.5pt minus1.5pt
                  \medskipamount=11.25pt plus3pt minus3pt
                  \bigskipamount=22.5pt plus6pt minus6pt
                  \normalbaselineskip=22.5pt plus0pt minus0pt
                  \normallineskip=1pt
                  \normallineskiplimit=0pt
                  \jot=5.625pt
                  {\def\smallskip {\vskip\smallskipamount}}
                  {\def\medskip   {\vskip\medskipamount}}
                  {\def\bigskip   {\vskip\bigskipamount}}
                  {\setbox\strutbox=\hbox{\vrule 
                    height15.75pt depth6.75pt width 0pt}}
                  \parskip 11.25pt
                  \normalbaselines}
\def\doublespace {\smallskipamount=7.5pt plus2pt minus2pt
                  \medskipamount=15pt plus4pt minus4pt
                  \bigskipamount=30pt plus8pt minus8pt
                  \normalbaselineskip=30pt plus0pt minus0pt
                  \normallineskip=2pt
                  \normallineskiplimit=0pt
                  \jot=7.5pt
                  {\def\smallskip {\vskip\smallskipamount}}
                  {\def\medskip   {\vskip\medskipamount}}
                  {\def\bigskip   {\vskip\bigskipamount}}
                  {\setbox\strutbox=\hbox{\vrule 
                    height21.0pt depth9.0pt width 0pt}}
                  \parskip 15.0pt
                  \normalbaselines}
\def\be{\begin{equation}}
\def\ee{\end{equation}}
\def\bea{\begin{eqnarray}}
\def\eea{\end{eqnarray}}

\def\sect #1{\setcounter{equation}{0}}

\begin{document}

\author{Sukratu Barve$^{*a}$, T. P. Singh$^{*b}$ and Louis Witten$^{\#c}$ 
\\\bigskip%
\  \\ %EndAName
$^{*}$Tata Institute of Fundamental Research,\\ Homi Bhabha Road, Mumbai
400005, India\\$^a$e-mail address: sukkoo@relativity.tifr.res.in\\$^b$e-mail
address: tpsingh@tifr.res.in\\\bigskip\ \\$^{\#}$Department of Physics,
University of Cincinnati,\\ Cincinnati, OH 45221-0011, USA\\$^c$e-mail
address: witten@physics.uc.edu}
\title{Spherical Gravitational Collapse: Tangential Pressure and Related 
Equations of State}
\date{}
\maketitle

\begin{abstract}
\noindent We derive an equation for the acceleration of a fluid element in
the spherical gravitational collapse of a bounded compact object made up of
an imperfect fluid. We show that non-singular as well as singular solutions
arise in the collapse of a fluid initially at rest and having only a 
tangential pressure. We obtain an exact solution of Einstein equations, in the
form of an infinite series, for collapse under tangential pressure with a 
linear equation of state. We show that if a singularity forms in the 
tangential pressure model, the conditions for the singularity to be naked are 
exactly the same as in the model of dust collapse.
\end{abstract}

\newpage\ \middlespace

\section{Introduction}

The study of gravitational collapse of a bounded spherical object in
classical general relativity has received major attention over the last few
years. The purpose of these investigations has been twofold - to establish
whether or not naked singularities arise in gravitational collapse \cite
{revi1}, and to study the occurrence of critical phenomena in
gravitational collapse \cite{revi2}.

Various models of spherical collapse have been studied over the last few
years, and these show that both black holes and naked singularities arise
during gravitational collapse. The collapsing matter is assumed to satisfy
one or more energy conditions, and is in this sense regarded as physically
reasonable matter. The models studied so far include collapse of dust \cite
{dust}, null dust \cite{vai}, perfect \cite{pf} and imperfect fluids \cite
{ipf}, and scalar fields \cite{ch}. In each of these cases, the formation of
covered as well as naked singularities has been observed. There are also
demonstrations that both these types of solutions will arise in the collapse
of a general form of matter \cite{gen}. Also, a naturalness argument has
recently been put forth to suggest that covered as well as naked singular
solutions arise generically in spherical collapse, subject to the assumption
of the dominant energy condition \cite{nat}.

Critical behaviour has been discovered in the collapse of matter fields and
fluids, largely through numerical studies. In some of these studies, it has
been observed that the solution separating dispersive solutions from
collapsing ones is a naked singularity. It should however be noted that in
studies of critical behaviour, the numerical identification of a ``black
hole'' is carried out by the detection of an apparent horizon. The
possibility that some of these ``black hole'' solutions are actually naked
singularities cannot be a priori ruled out. If the singularity is globally
naked, there would be a Cauchy horizon lying outside the apparent horizon,
in the Penrose diagram. Such a singularity cannot be identified in present
numerical studies, which do not probe the central high curvature region of
the collapsing object. On the other hand, a few analytical studies of scalar
collapse confirm that the non-dispersive solutions contain both black holes
and naked singular solutions \cite{rob}.

In the present paper, we study the spherical gravitational collapse of an
imperfect fluid under the assumption that the radial pressure is identically
zero, but the tangential pressure is non-zero. This system has been studied
by a few authors in the past \cite{tan1}, and also more recently \cite{tan2}%
, with the recent studies focusing on the issue of naked singularity
formation. The works mentioned in \cite{tan2} give evidence for naked
singularity formation. The purpose of the present paper is to demonstrate
the occurrence of non-singular and covered and naked singular solutions in a
specific tangential pressure model. As we will show, our analytical results
are similar, in some respects, to numerical results of critical behaviour in
collapse. We consider an equation of state for the tangential pressure $p_T,$
of the form $p_T=k(r)\rho $, for which we obtain an exact solution of the
Einstein equations, in the form of an infinite series. For physical reasons, we
assume that the function $k(r)$ must vanish at the origin of coordinates. As a
consequence of this constraint we find that if a singularity forms in this
model (which is not always the case) the conditions for the occurrence of a
naked singularity are exactly the same as in the dust collapse model.

The focus of our investigation will be the singularity that may possibly
form during the collapse, at the origin $r=0$ of the spherical coordinates.
This is the so-called central shell-focusing singularity. Hence we will
specify initial conditions only in a small neighborhood of the center and
investigate the nature of the collapse in that region, without considering
the evolution in the other regions of the spherical object. We will also
assume that the initial conditions are such that shell-crossing
singularities do not form during the evolution.

In Section 2 of the paper we write down the Einstein equations for the
collapse of a bounded spherical object in an asymptotically flat spacetime.
We then derive the equation for the acceleration of a fluid element in this
spherical object, under the mutual influence of gravity and the two
pressures (radial and tangential). In Section 3 we use the acceleration
equation to derive sufficient conditions for singularity formation, for the
cases of collapse under tangential pressure, collapse under radial pressure
and collapse of a perfect fluid. In all these cases we show that there are
initial conditions for which the evolution is non-singular, and other
initial conditions which result in singularity formation. In Section 4 we
give an exact series solution for the evolution of the area radius, in the
tangential pressure model under consideration. We show that the same
solution also holds for a very specific kind of evolution under radial
pressure or for a perfect fluid. In Section 5 we give a simplified
derivation of the earlier results on naked singularity formation in the dust
model. In Section 6 we show that covered as well as naked singularities form
in the tangential pressure model as well as the radial pressure and perfect
fluid models.

\section{The Acceleration Equation}

In comoving coordinates $(t,r,\theta ,\phi )$ the spherically symmetric
line-element is given by

\begin{equation}
\label{metric}ds^2=e^\sigma dt^2-e^\omega dr^2-R^2d\Omega ^2 
\end{equation}
where $\sigma $ and $\omega $ are functions of $t$ and $r$. The area radius $%
R$ also depends on both $t$ and $r.$ In comoving coordinates the
energy-momentum tensor for a spherically symmetric object takes the diagonal
form $T_k^i=(\rho ,-p_r,-p_T,-p_T)$. The quantities $p_r$ and $p_T$ are
interpreted to be the radial and tangential pressure, respectively. The
Einstein field equations for this system are

\begin{equation}
\label{mprime}m^{\prime }=4\pi \rho R^2R^{\prime }, 
\end{equation}

\begin{equation}
\label{mdot}\dot m=-4\pi p_rR^2\dot R, 
\end{equation}

\begin{equation}
\label{sigpri}\sigma ^{\prime }=-\frac{2p_r^{\prime }}{\rho +p_r}+\frac{%
4R^{\prime }}{R(\rho +p_r)}(p_T-p_r), 
\end{equation}

\begin{equation}
\label{omedot}\dot \omega =-\frac{2\dot \rho }{\rho +p_r}-\frac{4\dot R(\rho
+p_T)}{R(\rho +p_r)}, 
\end{equation}
and

\begin{equation}
\label{energy}m=\frac 12R\left( 1+e^{-\sigma }\dot R^2-e^{-\omega }R^{\prime
^2}\right) . 
\end{equation}
Here, $m(t,r)$ is a free function arising out of integration of the Einstein
equations. Its initial value, $m(0,r),$ is interpreted as the mass interior
to the coordinate $r$.

In order to derive the equation for the acceleration we first define $%
e^{-\omega }R^{\prime ^2}=1+f(t,r)$. Then

\begin{equation}
\label{fdot}\frac{\dot f}{1+f}=-\dot \omega +\frac{2\dot R^{\prime }}{%
R^{\prime }}=\frac{\dot R}{R^{\prime }}\sigma ^{\prime } 
\end{equation}
where the last equality follows from using (\ref{omedot}) and eliminating $%
\dot \rho $ using (\ref{mprime}) and (\ref{mdot}). Now we differentiate (\ref
{energy}) w.r.t. $t$ after writing it as 
\begin{equation}
\label{en2}e^{-\sigma }\dot R^2=\frac{2m}R+f 
\end{equation}
which gives the acceleration equation

\begin{equation}
\label{accn}\ddot R=-e^\sigma R\left( 4\pi p_r+\frac m{R^3}\right) +\frac
12\dot \sigma \dot R+\frac{e^\sigma (1+f)}{2R^{\prime }}\left( -\frac{%
2p_r^{\prime }}{\rho +p_r}+\frac{4R^{\prime }(p_T-p_r)}{R(\rho +p_r)}\right)
. 
\end{equation}
By defining the proper time $d\tau =e^{\sigma /2}dt$ this equation can also
be written as 
\begin{equation}
\label{accn2}\frac{d^2R}{d\tau ^2}=-R\left( 4\pi p_r+\frac m{R^3}\right) +%
\frac{(1+f)}{2R^{\prime }}\left( -\frac{2p_r^{\prime }}{\rho +p_r}+\frac{%
4R^{\prime }(p_T-p_r)}{R(\rho +p_r)}\right) 
\end{equation}
For a perfect fluid, we have $p_T=p_r\equiv p,$ and the above equation
becomes (see for instance \cite{ms}), 
\begin{equation}
\label{ms}\frac{d^2R}{d\tau ^2}=-R\left( 4\pi p+\frac m{R^3}\right) -\frac{%
(1+f)}{R^{\prime }}\frac{p^{\prime }}{\rho +p}. 
\end{equation}
The Oppenheimer-Volkoff equation for hydrostatic equilibrium is obtained by
setting the acceleration and the velocity equal to zero, and by noting, from
(\ref{en2}), that $f=-2m/R$: 
\begin{equation}
\label{ov}-R^2\frac{dp}{dR}=m\rho \left[ 1+\frac p\rho \right] \left[ 1+%
\frac{4\pi R^3p}m\right] \left[ 1-\frac{2m}R\right] . 
\end{equation}
A few interesting properties about the role of pressure in the acceleration
equation (\ref{accn}) should be noted. The tangential pressure appears only
in the last term, and its gradient does not enter the equation. The gradient
of only the radial pressure appears in the equation. A positive tangential
pressure opposes collapse, while a negative tangential pressure supports it.

\section{Conditions for singularity formation}

There are various interesting special cases of the Einstein equations for
spherical collapse given in the previous section, and we consider them one
by one. The dust approximation is obtained by setting $p_r=p_T=0.$ In this
case equation (\ref{mprime}) remains as such, while equation (\ref{mdot})
gives that the mass function does not depend on time $t$. Equation (\ref
{sigpri}) implies that $\sigma $ is a function only of time; hence we can
redefine $t$ and set $\sigma =0$. Equation (\ref{omedot}) can be integrated
to get $e^\omega =R^{\prime 2}/(1+f(r))$, where $f(r)$ is a function of
integration. Hence (\ref{energy}) can be written as 
\begin{equation}
\label{de}\dot R^2=\frac{2m(r)}R+f(r). 
\end{equation}
The dust model has been discussed in detail by many authors \cite{dust}.

\subsection{ Tangential Pressure}

In the case when the radial pressure $p_r$ is zero, and the tangential
pressure non-zero, considerable simplification of the full system of
equations takes place. As a result of equation (\ref{mdot}), the mass
function $m(r)$ is time-independent. Equations (\ref{sigpri}) and (\ref
{omedot}) become 
\begin{equation}
\label{st}\sigma ^{\prime }=\frac{4R^{\prime }}R\frac{p_T}\rho , 
\end{equation}
\begin{equation}
\label{ot}\dot \omega =-\frac{2\dot \rho }\rho -\frac{4\dot R}R\left( 1+%
\frac{p_T}\rho \right) . 
\end{equation}
The equation (\ref{accn}) for acceleration becomes 
\begin{equation}
\label{act}\ddot R=-e^\sigma \frac m{R^2}+\frac 12\dot \sigma \dot R+\frac{%
2e^\sigma (1+f)}R\frac{p_T}\rho . 
\end{equation}
We will assume that collapse begins from rest at time $t=0$, and choose the
initial scaling $R(0,r)=r,$ which gives from equation (\ref{en2}) that
initially $f=-2m(r)/r.$ In order for collapse to begin, the acceleration
must be negative, which implies that 
\begin{equation}
\label{cond}\frac{2m}R>\frac{4p_T/\rho }{1+4p_T/\rho }. 
\end{equation}

Collapse will continue all the way up to the formation of a singularity $R=0$%
, provided at any successive stage in the evolution, the acceleration is
negative when $\dot R=0$, i.e. provided (\ref{cond}) holds at all later
times. The ratio $p_T/\rho $ will in general evolve with time, for a given $r
$. Let the initial value of this ratio be denoted by $k(r)$, assumed to be
positive. A sufficient (though not necessary) condition for continual
collapse is that the ratio $p_T/\rho $ remains the same as, or falls below
its initial value $k(r).$ If this happens, then (\ref{cond}) will be
satisfied whenever $\dot R=0,$ because 
\begin{equation}
\label{cond3}\frac{2m(r)}R>\frac{2m(r)}r>\frac{4k(r)}{1+4k(r)}\geq \frac{%
4p_T/\rho }{1+4p_T/\rho }.
\end{equation}
We now examine collapse with the assumption that the ratio $p_T/\rho $
remains constant during evolution, at its initial value $k(r)$. This
assumption makes the analysis tractable. The constant $k(r)$ is chosen to
lie in the range $0<k(r)\leq 1$. Since the tangential pressure must vanish
at the origin as a result of isotropy, we must have $k(0)=0$. Equation (\ref
{omedot}) has the solution

\begin{equation}
\label{omesol}e^{-\omega (t,r)}=\chi (r)\rho ^2R^{4(1+k)}. 
\end{equation}
Here, $\chi (r)$ is an arbitrary function of the coordinate $r$ . Using
equation (\ref{mprime}) we can write this as

\begin{equation}
\label{omesol2}e^{-\omega (t,r)}=\frac{\chi (r)m^{\prime ^2}}{16\pi ^2}\frac{%
R^{4k}}{R^{\prime ^2}}\equiv C(r)\frac{R^{4k}}{R^{\prime ^2}}=\frac{1+f(t,r)%
}{R^{\prime ^2}}. 
\end{equation}
Note that $C(r)$ is a positive function. Using the solution (\ref{omesol2})
in equation (\ref{energy}) yields the following equation for the evolution
of the area radius: 
\begin{equation}
\label{reqn}\left( \frac{dR}{d\tau }\right) ^2=\frac{2m}R-1+C(r)R^{4k}. 
\end{equation}
Assuming that collapse begins from rest gives

\begin{equation}
\label{See}C(r)=\frac{\left( 1-\frac{2m}r\right) }{r^{4k}} 
\end{equation}
and hence (\ref{reqn}) becomes 
\begin{equation}
\label{reqnt}\left( \frac{dR}{d\tau }\right) ^2=\frac{2m}R-1+\left( 1-\frac{%
2m}r\right) \left( \frac Rr\right) ^{4k(r)}. 
\end{equation}

Let us consider the condition (\ref{cond}) for continual collapse. We are
interested in the behaviour near the origin, $r=0$. Near the origin, let $%
k(r)$ behave as a power law, $k(r)=ar^n$. Since $2m/r$ goes as $r^2$ near
the origin, collapse cannot take place if $n=1$, and will necessarily take
place if $n\geq 3$. The case $n=2$ is critical. Now, collapse will take
place provided $a<2\pi \rho _0/3$, but not otherwise. Here, $\rho _0$ is the
initial central density. Thus we find that for positive $k$ there are
singular as well as non-singular solutions. In Section 5 we will discuss
conditions for the singularities to be covered or naked.

\subsection{Perfect Fluid}

The perfect fluid approximation is obtained by setting $p_r=p_T=p$.
Equations (\ref{mprime}), (\ref{mdot}) and (\ref{energy}) remain as such,
while equations (\ref{sigpri}) and (\ref{omedot}) become 
\begin{equation}
\label{spf}\sigma ^{\prime }=-\frac{2p^{\prime }}{\rho +p}, 
\end{equation}
\begin{equation}
\label{opf}\dot \omega =-\frac{2\dot \rho }{\rho +p}-\frac{4\dot R}R. 
\end{equation}
The acceleration equation becomes 
\begin{equation}
\label{apf}\frac{d^2R}{d\tau ^2}=-R\left( 4\pi p+\frac m{R^3}\right) -\frac{%
(1+f)}{R^{\prime }}\frac{p^{\prime }}{\rho +p}. 
\end{equation}
For collapse starting from rest at time $t=0$ and by using the initial
scaling $R=r$ we get that initially $1+f=1-2m(r)/r$. In order for the
acceleration to be negative initially, we need 
\begin{equation}
\label{apn}\frac{2m}r>\frac{1+4\pi pr(\rho +p)/p^{\prime }}{1-(\rho
+p)/2rp^{\prime }}. 
\end{equation}
Let the initial density profile near the center be 
\begin{equation}
\label{dp}\rho =\rho _0-\alpha r^n,\;\alpha >0,\;n\geq 2 
\end{equation}
and let the equation of state be $p=k\rho $, with $k$ a positive constant.
For $n=2$ the condition for negative initial acceleration is 
\begin{equation}
\label{ac}\alpha <2\pi (k+1)\rho _0\left( 1+\rho _0/3k\right) . 
\end{equation}

For $n\geq 3$ the initial acceleration is necessarily negative. It can be
checked that a sufficient condition for singularity formation is that the
quantity $pR^2$ increases over, or remains the same as its initial value,
and further, the quantity $R|dp/dR|/(\rho +p)$ decreases from, or remains
the same as, its initial value.

\subsection{Radial Pressure}

Corresponding equations may be written for the case when the tangential
pressure $p_T$ is zero, and collapse takes place only under radial pressure.
It can be shown that starting from rest, the fluid will undergo collapse
leading to singularity formation, provided the following condition is
satisfied initially and subsequently: 
\begin{equation}
\label{rp}\frac{2m}R>\frac{8\pi p_rR^2+\left[ 2p_r^{\prime }R+4p_rR^{\prime
}\right] \,/\,(\rho +p_r)R^{\prime }}{-1+\left[ 2p_r^{\prime
}R+4p_rR^{\prime }\right] \,/\,(\rho +p_r)R^{\prime }}. 
\end{equation}
Consider the initial density profile near the center to be of the form (\ref
{dp}) (now with $n\geq 1$), and an equation of state $p_r=k(r)\rho $. Since
the radial pressure must vanish at $r=0$, $k(r)$ must vanish at the origin.
Let $k(r)=Ar^n$ near the center. Then it follows from equation (\ref{rp})
that collapse will not take place for $n=1$, will necessarily take place for 
$n\geq 3$, and if $n=2$, collapse will take place provided $\rho _0>\pi /6A$%
. A sufficient condition for singularity formation is that the quantity on
the right hand side of (\ref{rp}) remains constant or falls below its
initial value.

\section{An exact solution}

We will be interested in solving the collapse equation (\ref{reqnt}), which
describes the collapse of a fluid starting from rest, subject to a
tangential pressure equation of state $p_T=k(r)\rho $, with $k(0)=0$. If
this equation can be solved for a given $m(r)$, the remaining unknown
functions ($\rho ,\omega ,\sigma $ and $p_T$) can also be obtained.

Interestingly enough, there is also a class of evolutions for the perfect
fluid case and the radial pressure case, for which the equation for
evolution of the area radius can be cast in a form similar to (\ref{reqnt}).
Consider evolutions of the kind 
\begin{equation}
\label{gass}8\pi pR^2=\theta (r),\;\sigma ^{\prime }=\psi (r)\frac{R^{\prime
}}R. 
\end{equation}

One cannot attach any physical importance to these assumptions, and they may
at best only be approximately obeyed during the evolution. Their advantage
is that one can then obtain a solution, subject to these assumptions, and
demonstrate critical behaviour, and the occurrence of covered as well as
naked singularities, in this solution. Our main purpose is to give a
solution for the tangential pressure problem. It so happens that on the
side, we can say something useful about a specific perfect fluid model,
whose evolution equations are similar to the tangential pressure model.
However, the fact that an exact solution is possible could be a motivation
for understanding the assumptions on physical grounds. We would like to
point out the following in that context.

It can be easily seen from equation (\ref{metric}) that $\sigma $ plays the
role of the Newtonian potential in the weak field limit, suggesting that $%
\sigma ^{\prime }$ represents the gravitational pull on the source matter.
One could imagine a fluid source for the Einstein field equations wherein
the fluid particles were moving as shells of constant $r$. The mean
curvature of these shells would then be $R^{\prime }/R$. Also, $8\pi pR^2$
is analogous to the force experienced by the body of fluid enclosed by the
shell, solely due to the dynamics of the source particles.

With the assumptions made in (\ref{gass}) we can integrate equation (\ref
{mdot}) to get 
\begin{equation}
\label{mint}2m(t,r)=\theta (r)(r-R)+2m_0(r) 
\end{equation}
where $m_0(r)$ is the initial mass distribution. Similarly, (\ref{fdot}) and
(\ref{gass}) can be used to get 
\begin{equation}
\label{fpint}1+f(t,r)=A(r)R^{\psi (r)}, 
\end{equation}
where $A(r)$ is an integration constant. Assuming that collapse begins from
rest at $t=0$ (where the scaling is $R=r$), we get that $A(r)=(1-2m_0(r)/r)%
\,r^{-\psi }$.

To rewrite equation (\ref{energy}), the following transformation is made use
of:
\begin{equation}
\label{taudef}d\tau =e^{\sigma /2}dt+Z(r)dr 
\end{equation}
where the requirement of exactness of the differential equation restricts
the choice of $Z(r)$ to 
\begin{equation}
\label{Zeqn}Z(r)=\int e^{\sigma /2}\frac{\sigma ^{\prime }}2+g(r) 
\end{equation}
$g$ being an arbitrary function of $r$. We aim at solving equation (\ref
{energy}) keeping $r$ fixed. We vary $\tau $ under this restriction and
examining equation (\ref{energy}) obtain 
\begin{equation}
\label{pfm}\left( \frac{dR}{d\tau }\right) ^2=\frac{2m_0}R-1+\theta
(r)\left( \frac rR-1\right) -\left( \frac{2m_0(r)}r-1\right) \left( \frac
Rr\right) ^{\psi (r)}. 
\end{equation}

In case of a perfect fluid, the assumptions in (\ref{gass}) reduce to 
\begin{equation}
\label{ass}8\pi pR^2=\theta (r),\;-\;\frac{2Rdp/dr}{R^{\prime }(p+\rho )}%
=\psi (r). 
\end{equation}

For an equation of state $p=k\rho $ and an initial density profile given by (%
\ref{dp}) near the center, we get that near $r=0$, 
\begin{equation}
\label{sip}\psi (r)\sim \frac{n\alpha k}{(k+1)\rho _0}r^n. 
\end{equation}
After solving (\ref{pfm}) for $R(r,\tau )$ we can obtain the solutions for ($%
\rho ,\omega ,\sigma $ and $p_R).$

For collapse under radial pressure the assumptions (\ref{gass}) reduce to 
\begin{equation}
\label{ass2}8\pi pR^2=\theta (r),\;\;-\frac{4p_r^{\prime }R+8p_rR^{\prime }}{%
R^{\prime }(p+\rho )}=\psi (r). 
\end{equation}
For an equation of state of the form $p_r=k(r)\rho $, with $k(r)=A_0r^n$, we
get that $\psi (r)=-4(n+2)k(r)$.

The equation (\ref{reqnt}) for evolution of the area radius in the case of
collapse with tangential pressure is a special case of (\ref{pfm}), obtained
by setting $\theta (r)=0$ and $\psi (r)=4k(r)$. We now obtain an exact
series solution of Equation (\ref{pfm}), for a general $\theta (r)$ and $%
\psi (r)$. After defining the variable $y=R/r$ we write (\ref{pfm}) as an
integral: 
\begin{equation}
\label{bigint}\int d\tau =-\int \frac{r\sqrt{y}dy}{\sqrt{\theta +2m_0/r}}%
\left( 1-\frac{1+\theta }{\theta +2m_0/r}y+\frac{1-2m_0/r}{\theta +2m_0/r}%
y^{1+\psi }\right) ^{-1/2}. 
\end{equation}
By using the binomial expansion 
\begin{equation}
\label{bin}(1-x)^{-1/2}=1+\sum_{n=1}^\infty \frac{1.3.5...(2n-1)}{2^n\,n!}%
x^n 
\end{equation}
and by taking 
\begin{equation}
\label{exx}x=\frac{1+\theta }{\theta +2m_0/r}y+\frac{2m_0/r-1}{\theta +2m_0/r%
}y^{1+\psi } 
\end{equation}
the integral becomes 
\begin{equation}
\label{in2}\int d\tau =-\frac r{\sqrt{\theta +2m_0/r}}\left[ \frac
23y^{3/2}+\sum_{n=1}^\infty \frac{1.3.5...(2n-1)}{2^n\,n!}\int x^n\sqrt{y}%
dy\right] . 
\end{equation}
We write $x^n$ as 
\begin{equation}
\label{xn}x^n=\frac{y^n}{(\theta +2m_0/r)^n}\sum_{j=0}^{n}\,^nC_j\,(1+\theta
)^{n-j}\left( \frac{2m_0}r-1\right) ^jy^{j\psi (r)} 
\end{equation}
and hence get the solution to the integral as 
\begin{equation}
\nonumber
\tau -\tau _0(r)=-\frac {r}{\sqrt{\theta +2m_0/r}}\sum_{n=0}^{\infty} \frac{%
(2n)!}{2^{2n}(n!)^{2}}\frac {1}{(\theta +2m_0/r)^{n}}\times 
\end{equation}

\begin{equation}
\label{sol}\qquad \qquad \qquad \sum_{j=0}^n \,^nC_j\,(1+\theta
)^{n-j}\left( \frac{2m_0}r-1\right) ^j\frac{y^{n+j\psi +3/2}}{n+j\psi +3/2}. 
\end{equation}
This is an exact solution of the Einstein equation (\ref{pfm}).

We cast this into the following form for convenience, 
\begin{equation}
\label{solform}\tau -\tau _0(r)=-\frac r{\sqrt{\theta +2m_0/r}}y^{3/2}G(y,r) 
\end{equation}
where

$$
G(y,r)= \sum_{n=0}^{\infty} \frac{(2n)\!}{2^{2n}(n\!)^2} \frac {1}{(\theta
+2m_0/r)^{n}} \times 
$$

\begin{equation}
\label{Gdef}\qquad \qquad \qquad \sum_{j=0}^n\,^nC_j\,(1+\theta
)^{n-j}\left( \frac{2m_0}r-1\right) ^j\frac{y^{n+j\psi +3/2}}{n+j\psi }. 
\end{equation}

It is useful to work out the derivatives of $G$ : 
\begin{equation}
\label{Gydef}\left( \frac{\partial G}{\partial y}\right) _r=-\frac 32\frac{%
G(y,r)}y+\frac 1y\sqrt{\frac{2m_0/r+\theta -(1+\theta )y-(2m_0-1)y^{\psi +1}%
}{2m_0/r+\theta },} 
\end{equation}

$$
\left( \frac{\partial G}{\partial r}\right) _y=\sum_{n=0}^\infty \frac{(2n)!%
}{2^{2n}(n!)^2}\frac 1{(\theta +2m_0/r)^n}\times 
$$
\begin{equation}
\label{Grdef}\qquad \qquad \qquad \qquad \qquad \qquad
\sum_{j=0}^n\,^nC_j\,(1+\theta )^{n-j}\left( \frac{2m_0}r-1\right) ^j\frac{%
y^{n+j\psi }D(y,r,n,j)}{n+j\psi +3/2} 
\end{equation}
where 
$$
D(y,r,n,j)=(n-j)\frac{\theta ^{\prime }}{1+\theta }+j\frac{2m_0^{\prime
}/r-2m_0/r^2}{2m_0/r-1}-n\frac{2m_0^{\prime }/r-2m_0/r^2+\theta ^{\prime }}{%
2m_0/r+\theta }+ 
$$
\begin{equation}
\label{Ddef}\qquad \qquad \qquad j\psi ^{\prime }\log (y)-\frac{j\psi
^{\prime }}{n+j\psi +3/2}. 
\end{equation}
For the special case of dust, we have $\theta =\psi =0$, and $\tau =t$. It
can easily be shown from the dust equation (\ref{de}) for the case of
collapse starting from rest that the solution is given by 
\begin{equation}
\label{dsol}t-t_0(r)=-\frac{R^{3/2}}{\sqrt{2m(r)}}\left[ \frac{\arcsin \sqrt{%
y}}{y^{3/2}}-\frac{\sqrt{1-y}}y\right] . 
\end{equation}
The series solution (\ref{sol}) can be summed up in the dust case and it can
be shown that the sum is equal to the closed form solution given in (\ref
{dsol}).

\section{The Dust Solution}

We will be interested in using the solution (\ref{sol}) to find out whether
the singularity that forms at $r=0$ is covered or naked. We start by noting
that in the cases of interest discussed above, $\psi (r)$ has a power law
form, near $r=0$. Also, $\theta (r)$ has a power law form near $r=0$.
Because of this, it can be inferred from (\ref{sol}) that as $r\rightarrow 0$
the solution approaches the dust solution given in (\ref{dsol}), with the
difference that in (\ref{dsol}), $t$ should be replaced by $\tau $ and $%
2m(r) $ should be replaced by $2m_0(r)+r\theta (r)$, during the approach.

Hence it should be possible, as explicitly shown later, to use the naked
singularity analysis for dust, carried out earlier, to draw conclusions
about the occurrence of a naked singularity in collapse of a fluid under
tangential pressure or radial pressure, or a perfect fluid, subject to the
equation of state chosen above. Before we do so, we would like to use this
opportunity to present a simplified derivation of the dust naked
singularity. We give this simplified analysis first for the marginally bound
case $f(r)=0$, and then for the general case.

The solution to equation (\ref{de}) in the marginally bound case is 
\begin{equation}
\label{ds}R^{3/2}=r^{3/2}-\frac 32\sqrt{2m(r)}t. 
\end{equation}
An initial scaling $R=r$ at the starting epoch $t=0$ of the collapse has
been assumed. A curvature singularity forms at $r=0$ at time $t_0=2/3\sqrt{%
\rho _0}$, where $\rho _0$ is the initial central density. We assume a
series expansion near $r=0$ for the initial density function $\rho (r)$, 
\begin{equation}
\label{ind}\rho (r)=\rho _0+\rho _1r+\frac 1{2!}\rho _2r^2+\frac 1{3!}\rho
_3r^3+... 
\end{equation}
The series expansion for the mass $m(r)$ can then be deduced using (\ref
{mprime}).

From equation (\ref{ds}), we evaluate $R^{\prime }$ and then substitute for $%
t$ from the same equation. In the resulting expression, substitute $%
R=Xr^\alpha $, and perform a Taylor expansion of $F(r)$ around $r=0$, so as
to retain the leading non-vanishing term. We then get that near $r=0$,

\begin{equation}
\label{rpr}\frac{R^{\prime }}{r^{\alpha -1}}=X+\frac \beta {\sqrt{X}%
}r^{q+3/2-3\alpha /2}. 
\end{equation}
Here, $\beta =-2qm_q/3m_0$, and $q$ is defined such that in a series
expansion of the initial density $\rho (r)$ near the center, the first
non-vanishing derivative is the $q$th one (=$\rho _q$), and $m_q=4\pi \rho
_q/(q+3)q!$. It may be shown that $\alpha$ has a unique value at the approach 
to the central singularity, given by setting the power of $r$ to zero in the
second term, i.e. $\alpha =1+2q/3$. This
reproduces the result of the $R^{\prime }$ calculation performed earlier 
\cite{sin}, in a simpler manner.

For the non-marginally bound case (i.e. $f\neq 0$) the solution of the 
Tolman-Bondi equation is

\begin{equation}
\label{nmar}R^{3/2}G(-fR/F)=r^{3/2}G(-fr/F)-\sqrt{F(r)}t 
\end{equation}
where $F(r)=2m(r)$, and $G(y)$ is a positive function having the domain $%
1\geq y\geq -\infty $ and is given by. $R^{\prime }$ can be evaluated as
before, and then we eliminate $t$ and substitute $R=Xr^\alpha $. The
power-series expansion for $f(r)$ near $r=0$ is of the form \cite{sin}

\begin{equation}
\label{fx}f(r)=f_2r^2+f_3r^3+f_4r^4+... 
\end{equation}
which implies that the argument ($-fR/F$) of $G$ on the left-side of ($\ref
{nmar})$ goes to zero as $r\rightarrow 0$. The derivative of $G(-p)$ w.r.t.
its argument $p\equiv rf/F$ is obtained by differentiating (\ref{nmar}),
which gives

$$
\frac{dG(-p)}{d(-p)}=\frac{3G}{2p}-\frac 1{p\sqrt{1+p}}. 
$$
Using this, one can now perform a Taylor-expansion of $G,F$ and $f$ about $%
r=0$ to get exactly the same expression for $R^{\prime }$ as given in (\ref
{rpr}) above, where $\beta $ is now given by

\begin{equation}
\label{the}\beta =q\left( 1-\frac{f_2}{2F_0}\right) \left( G(-f_2/F_0)\left[ 
\frac{F_q}{F_0}-\frac{3f_{q+2}}{2f_2}\right] \left[ 1+\frac{f_2}{2F_0}%
\right] +\frac{f_{q+2}}{f_2}-\frac{F_q}{F_0}\right) . 
\end{equation}
The constant $q$ is now defined such that the first non-vanishing derivative
of the initial density is the $q$th one, and/or the first non-vanishing term
in the expansion for $f(r)$, beyond the quadratic term, is of order $r^{q+2}$%
. The constant $\alpha $ is again equal to $(1+2q/3)$. Thus the $R^{\prime }$
calculation is simplified for the non-marginal case as well.

In order for the singularity at $r=0$ to be naked, radial null geodesics
should be able to propagate outwards, starting from the singularity. A
necessary and sufficient condition for this to happen is that the area
radius $R$ increase along an outgoing geodesic, because $R$ becomes negative
in the unphysical region. Thus we write, along the geodesic, 
\begin{equation}
\label{Rr}\frac{dR}{dr}=R^{\prime }+\dot R\frac{dt}{dr}=R^{\prime }\left( 1+%
\frac{\dot R}{\sqrt{1+f}}\right) =R^{\prime }\left( 1-\sqrt{\frac{f+F/R}{1+f}%
}\right) . 
\end{equation}
Here we have substituted $dt/dr=R^{\prime }/\sqrt{1+f}$ along an outgoing
null ray (using the metric (\ref{metric})) and substituted for $\dot R$ from
(\ref{de}). $dR/dr$ should be positive along the outgoing geodesic. We now
define $u=r^\alpha $, and use $X$ as a variable, instead of $R$. Hence, in
the approach to the singularity, (i.e. as $R\rightarrow 0,r\rightarrow 0)$, $%
X$ takes the limiting value $X_0$ given by 
\begin{equation}
\label{xz}X_0=\lim \limits_{r\rightarrow 0,R\rightarrow 0}\frac Ru=\lim
\frac 1{\alpha r^{\alpha -1}}\frac{dR}{dr}=\lim \frac 1{\alpha r^{\alpha
-1}}R^{\prime }\left( 1-\sqrt{\frac{f+F/R}{1+f}}\right) . 
\end{equation}
By using (\ref{rpr}) we can write 
\begin{equation}
\label{re}X_0=\frac 1\alpha \left( X_0+\frac \beta {\sqrt{X_0}}\right)
\left( 1-\sqrt{\frac{f(0)+\Lambda _0/X_0}{1+f(0)}}\right) 
\end{equation}
The constant $\Lambda _0$ is the limiting value of $\Lambda
(r)=F(r)/r^\alpha $ as $r\rightarrow 0$. The variable $X$ can be interpreted
as the tangent to the outgoing geodesic, in the $R,u$ plane. As can be seen
from the above, the positivity of $dR/dr$ along an outgoing geodesic is
equivalent to requiring that the equation (\ref{re}) admit a positive root $%
X_0$. This will depend on the initial density and velocity distribution,
which determine the functions $F(r)$ and $f(r)$, and hence the functions $%
\beta $ and $\Lambda $. One can solve equation (\ref{re}), and as shown in 
\cite{sin}, the results for the nature of the singularity are the following:

In the marginally bound case, the singularity is naked if $\rho _1<0$, or if 
$\rho _1=0,\rho _2<0.$ If $\rho _1=0$ and $\rho _2=0,$ one defines the
quantity $\zeta =2m_3/(2m_0)^{5/2}.$ The singularity is naked if $\zeta \leq
-25.9904$ and covered if $\zeta $ exceeds this value. If $\rho _1=$ $\rho
_2=\rho _3=0,$ the singularity is covered. In the non-marginally bound case,
if $F_1$ and $f_3$ are non-zero, the singularity is naked if 
\begin{equation}
\label{q1}Q_1=\left( 1-\frac{f_2}{2F_0}\right) \left( G(-f_2/F_0)\left[ 
\frac{F_1}{F_0}-\frac{3f_3}{2f_2}\right] \left[ 1+\frac{f_2}{2F_0}\right] +%
\frac{f_3}{f_2}-\frac{F_1}{F_0}\right) 
\end{equation}
is positive. If $F_1$ and $f_3$ are both zero, and $F_2$ and $f_4,$ are
non-zero the singularity is naked if 
\begin{equation}
\label{q2}Q_2=\left( 1-\frac{f_2}{2F_0}\right) \left( G(-f_2/F_0)\left[ 
\frac{F_2}{F_0}-\frac{3f_4}{2f_2}\right] \left[ 1+\frac{f_2}{2F_0}\right] +%
\frac{f_4}{f_2}-\frac{F_2}{F_0}\right) 
\end{equation}
is positive. If $F_1$, $f_3$, $F_2$ and $f_4$ are zero and if $F_3$ and $f_5$
are non-zero, the singularity is naked if 
\begin{equation}
\label{q3}Q_3=\left( 1-\frac{f_2}{2F_0}\right) \left( G(-f_2/F_0)\left[ 
\frac{F_3}{F_0}-\frac{3f_5}{2f_2}\right] \left[ 1+\frac{f_2}{2F_0}\right] +%
\frac{f_5}{f_2}-\frac{F_3}{F_0}\right) 
\end{equation}
is positive. If $F_1$, $f_3$, $F_2$, $f_4$, $F_3$ and $f_5$ are all zero,
the singularity is covered.

A special case of non-marginally bound collapse is the collapse starting
from rest, for which $f(r)=-2m(r)/r$. As a result, $G(-fR/F)=\pi /2$ and the
calculation of $R^{\prime }$ in (\ref{nmar}) can be carried out exactly as
in the marginally bound case, to get the result (\ref{rpr}), with $\beta
=-\pi qm_q/4m_0$. The conditions for a naked singularity to occur are the
same as those stated above for marginally bound collapse, except that $\zeta 
$ is defined as $\zeta =6\pi m_3/4(2m_0)^{5/2}$.

We also point out that in a recent work \cite{ba} we have given a yet
simpler derivation of the dust naked singularity, which directly looks for a
self-consistent solution of the geodesic equation, in a neighborhood of the
singularity. In principle, the method described in \cite{ba} can be applied
also to the tangential pressure solution discussed here.

\section{The occurrence of covered and naked singularities}

We can now utilize the dust results for inferring the occurrence of covered
and naked singularities in collapse with the equations of state considered
here, except that two further subtleties remain to be sorted out. Firstly,
the solution (\ref{sol}) is written with respect to the time variable $\tau $%
, for a {\it fixed} $r$. We need to show that a relation similar to equation
(\ref{xz}) holds. Secondly, for a general $\psi $ and $\theta $ (i.e. not
restricted to power law forms for instance) the dependence of $f$ and $F$ on 
$t$ as well introduces terms in $X_0$ different from the dust case. It is
not obvious a priori that these vanish as one approaches the singularity,
even though the solution tends to behave like dust in this limit. It needs
to be explicitly shown that a relationship identical in form to equation (%
\ref{re}) holds in the general case. This is demonstrated in the rest of the
section.

By requiring that $d\tau $ be integrable, we can write 
\begin{equation}
\label{int}d\tau =e^{\sigma /2}dt+Z(r)dr=e^{\sigma /2}dt+\left[ \int
e^{\sigma /2}\,\frac{\sigma ^{\prime }}2dt+g(r)\right] dr 
\end{equation}
where the last expression follows from using the integrability condition on $%
Z(r)$, and $g(r)$ is an arbitrary function. We choose $g(r)$ in such a way
that in the approach to the central singularity the coefficient of $dr$ in (%
\ref{int}) vanishes. As a result, in the approach to the singularity we can
write $d\tau =e^{\sigma /2}dt.$

The following can be easily shown from (\ref{pfm}) and the solution (\ref
{solform}). 
\begin{equation}
\label{long}\left( \frac{\partial R}{\partial \tau }\right) _r=-\frac 1{%
\sqrt{y}}\left[ A(y,r)\right] ^{1/2} 
\end{equation}
where 
\begin{equation}
\label{Adef}A(y,r)=\frac{2m_0(r)}r+\theta -(1+\theta )y-\left( \frac{2m_0}%
r-1\right) y^{1+\psi }. 
\end{equation}
$\partial R/\partial \tau $ is finite in the approach to the singularity.
Also,

\begin{equation}
\label{long'}\left( \frac{\partial R}{\partial r}\right) _\tau =y\left[
1+\frac r{\sqrt{2m_0/r+\theta }}\sqrt{A(y,r)}B(y,r)\right] 
\end{equation}
where 
\begin{equation}
\label{Bdef}B(y,r)=-\frac{\tau _0^{\prime }G}{\tau -\tau _0}-\frac Gr+1/2%
\frac{2m_0^{\prime }/r-2m_0/r^2+\theta ^{\prime }}{2m_0/r+\theta }-\left( 
\frac{\partial G}{\partial r}\right) _y. 
\end{equation}

Now, 
\begin{equation}
\label{ba}(R^{\prime })_t=(R^{\prime })_\tau +\left( \frac{\partial R}{%
\partial \tau }\right) _r\left( \frac{\partial \tau }{\partial r}\right) _t 
\end{equation}
and it follows from the above discussion that the last term, $M(y,r)$ say,
can be made to vanish in the approach to the singularity, so that in the
limit the quantities $(R^{\prime })_t$ and $(R^{\prime })_\tau $ become
identical to each other.

From assumptions (\ref{gass}) one can easily show that 
\begin{equation}
\label{eomega}R^{\prime }{}^2e^{-\omega }=(1-2m_0/r)\,y^\psi 
\end{equation}
This leads to 
\begin{equation}
\label{eomega/2}e^{\omega /2}=\frac{y^{1-\psi /2}}{\sqrt{1-2m_0/r}}\left[
1+\frac r{\sqrt{2m_0/r+\theta }}\sqrt{A(y,r)}B(y,r)\right] +\frac{y^{1-\psi
/2}}{\sqrt{1-2m_0/r}}M(y,r) 
\end{equation}

From (\ref{int}) we also get that along a null geodesic 
\begin{equation}
\label{tar}\frac{\partial \tau }{\partial r}=e^{\omega /2}+\int e^{\sigma
/2}\,\frac{\sigma ^{\prime }}2dt+g(r). 
\end{equation}
This, after using (\ref{eomega/2}), is re-written as 
$$
\frac{\partial \tau }{\partial r}=\frac{y^{1-\psi /2}}{\sqrt{1-2m_0/r}}%
\left[ 1+\frac r{\sqrt{2m_0/r+\theta }}\sqrt{A(y,r)}B(y,r)\right] +\frac{%
y^{1-\psi /2}}{\sqrt{1-2m_0/r}}M(y,r) 
$$
\begin{equation}
\label{tarexpl}\qquad \qquad \qquad +\int e^{\sigma /2}\,\frac{\sigma
^{\prime }}2dt+g(r) 
\end{equation}

Hence we may examine the rate of change of $R$ along an outcoming null ray
as 
\begin{equation}
\label{uh}\left( \frac{dR}{dr}\right) _{null\,geodesic}=\left( R^{\prime
}\right) _\tau +\left( \frac{\partial R}{\partial \tau }\right)
_{null\,geodesic}\left( \frac{\partial \tau }{\partial r}\right) 
\end{equation}
From equations (\ref{long}), (\ref{long'}) and (\ref{tarexpl}), this can be
cast as

$$
\left( \frac{dR}{du}\right) _{null\,geodesic}=\frac 1\alpha [X-\sqrt{\frac
A{\left( 2m_0/r+\theta \right) r^{3\left( \alpha -1\right) }}}\frac{\tau
_0^{\prime }}{\sqrt{X}}-\sqrt{\frac A{2m_0/r+\theta }}\sum_{n=0}^\infty 
\frac{(2n)\!}{2^{2n}(n\!)^2}\frac 1{(\theta +2m_0/r)^n} 
$$
$$
\qquad \sum_{j=0}^nX^{n+j\psi }\,^nC_j\,(1+\theta )^{n-j}\left( \frac{2m_0}%
r-1\right) ^j\frac{r^{\left( \alpha -1\right) \left( n+j\psi \right) }}{%
n+j\psi +3/2} 
$$
\begin{equation}
\label{huge}\qquad \left( D(y,r,n,j)+\frac 1r+1/2\frac{2m_0^{\prime
}/r-2m_0/r^2+\theta ^{\prime }}{2m_0/r+\theta }\right) ]\left( 1-X^{\left(
\psi +1\right) /2}\sqrt{\frac A{\left( 2m_0/r-1\right) \left( r^{\left(
\alpha -1\right) \left( n+j\psi \right) }\right) }}\right) 
\end{equation}
where the derivative on the left is to be evaluated along null geodesics.
This resembles the roots equation (\ref{Rr}).

Indeed, the first term with square parenthesis in equation (\ref{huge}) is
similar to $R^{\prime }/r^{\alpha -1}$ in equation (\ref{xz}). It turns out
that as one takes the limit $r\rightarrow 0$, the third contribution within
this term containing the summation vanishes for physically reasonable
choices of $\psi $, i.e. vanishes as one takes the limit. This requirement
is certainly satisfied by the three special cases in Sections 3.1, 3.2 and
3.3. In the rest of the contribution, these conditions imply that the result
is the same as dust, save the replacement of $2m$ by $2m_0+r\theta $ and $%
t_0 $ by $\tau _0$ in the approach to the singularity. Similarly, the second
term in square parenthesis in equation (\ref{huge}) has a counterpart in the
dust case, although the familiar $f$ and $F$ of (\ref{Rr}) have no
individual correspondences. However, if one examines the limiting form (\ref
{xz}) of dust and works out (\ref{huge}) in the limit, then one finds that $%
f(0)$ of (\ref{xz}) corresponds to $-2m_0/r-\theta $ evaluated in the limit $%
r=0$ and $\Lambda _0$ gets replaced by $(2m_0+r\theta )/r^\alpha $
(evaluated in the limit $r=0$). Hence we arrive at the same roots equation
as (\ref{re}).

It is important to realize that the first subtlety mentioned at the
beginning is sorted out by a judicious choice of the $\tau $ coordinate and
results in the fact that $(\partial R/\partial r)_\tau $ tends to $R^{\prime
}$ of the dust case in (\ref{Rr}) in the limit. The second subtlety presents
itself explicitly when it turns out that the correspondences of $f$ and $F$
of dust are not $f(r,t)$ and $F(r,t)$ of the general case if one has not
taken the limit. It would be therefore in general incorrect to conclude
equation (\ref{huge}) to be equation (\ref{Rr}) with $f$ and $F$ simply
replaced by their generalizations, at this stage. However, the term $%
(f+F/R)/(1+f)$ in (\ref{Rr}) still corresponds to the quantity obtained by
simple generalization of the free functions of dust. The existence of such a
quantity is not apriori guaranteed when one notices that the solution (\ref
{sol}) behaves similar to the dust solution in the approach to the
singularity.

We now discuss the nature of the singularities. Consider first collapse
under tangential pressure $p_T$ with the equation of state $p_T=k(r)\rho $,
which we have discussed above. Since in this case $\theta (r)=0$ it follows
from the exact solution given above in (\ref{sol}) that if the collapse ends
in a singularity, then the conditions for the occurrence of covered and
naked singularities are exactly the same as in the case of dust collapse
starting from rest. The introduction of a tangential pressure (which must
vanish at the center even though it is non-zero elsewhere) does not change
the nature of the dust singularity. Further, we expect that even if we do
not restrict to the case of linear equation of state, the conditions for a
singularity to be naked in the tangential pressure model will be exactly the
same as in the dust case, so long as the tangential pressure vanishes at the
origin.

The situation is more interesting in the perfect fluid case, because now $%
\theta (r)\neq 0$. The leading order solution is the dust solution, but with
an ``effective mass'' $2m_0(r)+\theta (r)$. Hence the dust results on the
nature of the singularity are applicable, except that we must replace $%
2m_0(r)$ by $2m_0(r)+\theta (r)$. Recalling the definition of $\theta (r)$,
we can expand it in a series near $r=0$, 
\begin{equation}
\label{thexp}\frac{\theta (r)}{4\pi }=p_0r^2+\frac 12p_2r^2+\frac
16p_3r^6+... 
\end{equation}
We have also assumed an equation of state $p=k\rho $, with $k$ a constant.
Hence the coefficients in the above expansion for the pressure are related
to those for the density in the expansion in (\ref{ind}). Eqns. (\ref{dp})
and (\ref{ac}) give the conditions for singularity formation, if $\rho _2<0$%
. If a singularity does form it will be naked. If $\rho _2=0$ and $\rho _3<0$
a singularity will necessarily form. It will be naked if $\zeta =3\pi
(2m_3+2\pi p_3/3)/4F_0^{5/2}$ is less than $-25.9904$ and covered if $\zeta $
exceeds this value. Thus we find that in the case of a perfect fluid, the
condition for the occurrence of a naked singularity differs from that in the
dust case, because of the presence of the constant $k$ in the definition for 
$\zeta $. By following a similar series of arguments one can conclude that
if a singularity forms in the radial pressure model considered above, the
conditions for it to be naked or covered are exactly the same as in the dust
case.

\section{ Appendix}

Although we have shown that the collapse of the cloud in Sections 3.1, 3.2
and 3.3 leads to the formation of a singularity for some initial conditions,
it is necessary to ensure that the same happens in case of the general
solution (\ref{sol}) obtained with the assumptions (\ref{gass}). We proceed
in a manner similar to these cases. Initially, let $\dot R$ be $0$ and $%
\ddot R$ be negative. If the collapse has to take place without any rebounds
hereafter, one needs to ensure that the shells have negative `acceleration' $%
\ddot R$ whenever they reach the velocity limit $\dot R=0$. This along with
the initial condition keeps $\dot R$ non-positive. It is straightforward to
show from assumptions (\ref{gass}) using (\ref{eomega}) and (\ref{energy})
that this requirement for avoiding rebounds reduces to the following
condition on $R$: 
$$
\frac Rr<\frac{\frac \psi 2\left( 2m_0/r+\theta \right) }{\left(
2m_0/r+\theta \right) +\left( 1+\theta \right) \frac \psi 2} 
$$
whenever $R$ satisfies 
$$
2m_0/r+\theta -(1+\theta )\frac Rr+(1-2m_0/r)\left( \frac Rr\right) ^{1+\psi
}=0. 
$$
Note that this is equivalent to an algebraic inequality dependent solely on
the initial data, when the root $R/r$ in the equation above is bounded by
the first inequality. Let us suppose this constraint on the initial free
functions is satisfied for one of the roots which we know to be $1$. This is
nothing but the initial epoch of the collapse. The collapse begins. Let us
suppose there occurs another real root (i.e. $\dot R$ vanishes again). This
time the collapse has proceeded and the root, therefore, has to be lesser
than $1$. The first inequality, already satisfied for $R/r=1$ will therefore
be automatically satisfied making the acceleration negative in this
situation. Thus the rebound will be prevented. Hence we conclude that if we
ensure that the initial data is chosen such that the cloud begins to
collapse at the first instant, then it is implied that the cloud will have
no rebounds at all at any later instant.

\bigskip\ 

{\noindent {\bf ACKNOWLEDGMENTS}}

\smallskip

\noindent We acknowledge partial support of the Junta Nacional de
Investigac\~ao Cient\'ifica e Tecnol\'ogica (JNICT) Portugal, under contract
number CERN/S/FAE/1172/97. L. W. acknowledges the partial support of NATO,
under contract number CRG 920096 and also the partial support of the U. S.
Department of Energy under contract number DOE-FG02-84ER40153.

\end{document}